\input{aipcheck.tex}

\documentclass{aipproc}
\usepackage{amssymb,amsmath}
\usepackage{euscript}
\usepackage{epsfig}

\def\selectedlayoutstyle{6x9}
\layoutstyle\selectedlayoutstyle

\SetInternalRegister\hbadness{8000} 

\newcommand\doingARLO[2][]{%
  \ifx\mmref\undefined #1\else #2\fi
}

\begin{document}

\newtheorem{theorem}{Theorem}
\newtheorem{definition}{Definition}

\title
      [Measuring Questions:\\
Relevance and its Relation to Entropy]
      {Measuring Questions:\\
Relevance and its Relation to Entropy}


\author{Kevin H. Knuth}
{address={Comp. Sci. Div., NASA Ames Research Ctr., M/S 269-3,
Moffett Field CA 94035}}

\copyrightyear{2004}

\begin{abstract}
The Boolean lattice of logical statements induces the free
distributive lattice of questions.  Inclusion on this lattice is
based on whether one question answers another. Generalizing the
zeta function of the question lattice leads to a valuation called
\textit{relevance} or \textit{bearing}, which is a measure of the
\textit{degree} to which one question answers another. Richard Cox
conjectured that this degree can be expressed as a generalized
entropy. With the assistance of yet another important result from
Janos Acz\'{e}l, I show that this is indeed the case, and that the
resulting inquiry calculus is a natural generalization of
information theory. This approach provides a new perspective on
the Principle of Maximum Entropy.
\end{abstract}

\date{\today}
\maketitle

\newcommand{\downset}[1]{\:\:\downarrow\!\!#1}

``A wise man's question contains half the answer.''~~Solomon Ibn
Gabirol (1021-1058)

\section{Questions and Answers}
Questions and answers, the unknown and the known, empty and full
are all examples of duality.  In this paper, I will show that a
precise understanding of the duality of questions and answers
allows one to determine the unique functional form of the
relevance measure on questions that is consistent with the
probability measure on the set of logical statements that form
their answers. Much of the material presented in this paper relies
on fundamental background material that I regrettably cannot take
the space to address. While I provide a brief background below, I
recommend the following previous papers
\cite{Knuth:Questions,Knuth:PhilTrans,Knuth:laws,Knuth:Neurocomp-preprint}
in which more background, along with useful references, can be
found.

\section{Lattices and Valuations}
A partially ordered set, or poset for short, is a set of elements
ordered according to a binary ordering relation, generically
written $\leq$. One element $b$ is said to `include' another
element $a$ when $a \leq b$. Inclusion on the poset is encoded by
the \emph{zeta function}
\begin{equation}
\label{eq:zeta} \zeta(x,y) =
   \left\{ \begin{array}{rl}
       1 & \mbox{if}~~x \leq y \\
       0 & \mbox{if}~~x \nleq y \end{array} \right.
   \mbox{~~~~~(\emph{zeta function})}
\end{equation}
If there is a greatest element in the poset, it is called the top
$\top$, and dually, there may be a bottom $\bot$. Given two
elements $a$ and $b$ of the poset, their upper bound is the set of
all elements $x$, such that $a \leq x$ and $b \leq x$, where $x
\neq a$ and $x \neq b$. If there exists a \textit{unique} least
upper bound, this element is called the \emph{join} of $a$ and
$b$, written $a \vee b$. Similarly, if there exists a greatest
lower bound, that element is called the \emph{meet}, written $a
\wedge b$. A \emph{lattice} is a poset in which unique joins and
meets of all pairs of elements exist. In this case, the join and
meet can be seen as binary operations that take two objects and
map them to a third. For this reason, lattices are algebras. When
one views the lattice as a set of elements ordered by an ordering
relation, one is taking a structural viewpoint. When it is viewed
as a set of elements and a set of operations, one is taking an
operational viewpoint, which is an algebra. Last, elements that
cannot be expressed as a join of two other elements are called
\emph{join-irreducible elements}.

Generalizing the zeta function allows one to define \emph{degrees
of inclusion}. In actuality, it is more useful to generalize the
dual of the zeta function \cite{Knuth:Neurocomp-preprint}, which
is the zeta function (\ref{eq:zeta}) with the conditions flipped
around. The result is a real-valued function that captures the
notion of degrees of inclusion

\begin{equation}
\label{eq:z} z(x, y) =
   \left\{ \begin{array}{rl}
       1 & \mbox{if}~~x \geq y \\
       0 & \mbox{if}~~x \wedge y = \bot\\
       z & \mbox{otherwise, where}~~0 < z < 1. \end{array} \right.
   \mbox{(\emph{degrees of inclusion})}
\end{equation}

The rules by which degrees of inclusion are manipulated as one
moves about the lattice are found by maintaining consistency with
the lattice structure, or equivalently the underlying algebra. All
lattices are associative, and as such, they all possess a sum
rule. With an important result from Caticha \cite{Caticha:1998}, I
have shown that all distributive lattices give rise
\cite{Knuth:laws,Knuth:Neurocomp-preprint} to a sum rule,
\begin{equation}
z(x_1 \vee x_2 \vee \cdots \vee x_n, t) = \sum_i{z(x_i, t)} -
\sum_{i<j}{z(x_i \wedge x_j, t)} + \sum_{i<j<k}{z(x_i \wedge x_j
\wedge x_k, t)} - \cdots \label{eq:incl-excl}
\end{equation}
a product rule
\begin{equation}
z(x \wedge y, t) = C z(x, t) z(y, x \wedge t),
\label{eq:product-rule}
\end{equation}
and a Bayes' Theorem
\begin{equation}
z(y, x \wedge t) = \frac{z(y, t) z(x, y \wedge t)}{z(x, t)}.
\label{BayesThm:appendix}
\end{equation}
This immediately conjures up thoughts of probability theory,
however this result is surprisingly far more general.

\section{Probability}
\begin{figure}[t]
\label{fig:fig1} \caption{The ordered set of down-sets of the
lattice of assertions $\EuScript{A}$ results in the corresponding
lattice of questions $\EuScript{Q} = \EuScript{O}(\EuScript{A})$
ordered by $\subseteq$. $\EuScript{A}$ is dual to $\EuScript{Q}$
in the sense of Birkhoff's Representation Theorem. The
join-irreducible elements of $\EuScript{Q}$ are the ideal
questions $\EuScript{I}$, which are isomorphic to the lattice
$\EuScript{A} \sim \EuScript{I} = \EuScript{J}(\EuScript{Q})$.
Down-sets corresponding to several questions are illustrated on
the right.}
\includegraphics[height=.34\textheight]{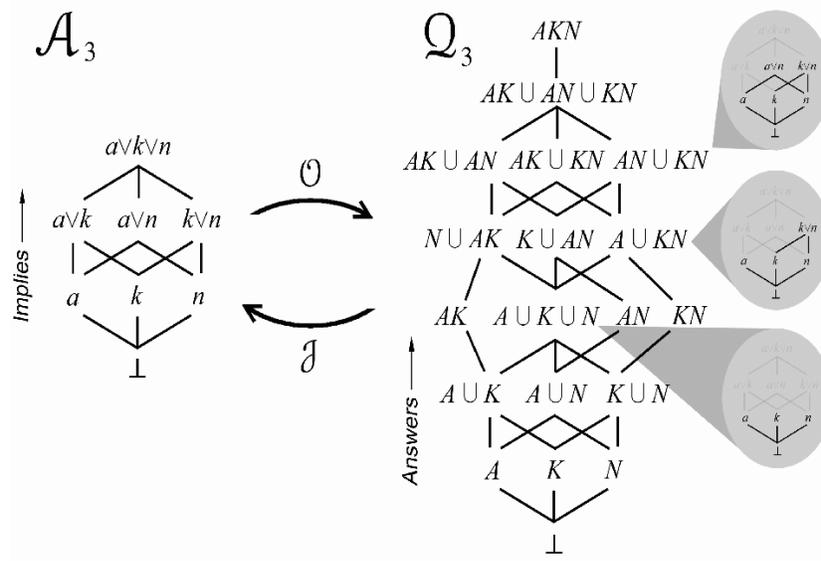}
\end{figure}

It is now well-understood that probability theory is literally an
extension of logic. A set of logical statements ordered by
implication gives rise to a Boolean lattice, which is equivalently
a Boolean algebra. Figure \ref{fig:fig1} shows the lattice
$\EuScript{A}_3$ generated from a set of three mutually exclusive
and exhaustive assertions: $a$, $k$, and $n$. This example is
taken from two previous papers, and deals with the issue of `Who
stole the tarts?' in Lewis Carroll's \textit{Alice in Wonderland},
specifically
\begin{align*}
a &= \textit{`Alice stole the tarts!'}\\
k &= \textit{`The Knave of Hearts stole the tarts!'}\\
n &= \textit{`No one stole the tarts!'}
\end{align*}
The lattice $\EuScript{A}_3$ shows all possible statements that
can be formed from these three atoms.

The zeta function of this lattice, which is a function of two
statements, indicates whether one statement implies another.
Generalizing the dual of the zeta function results in a
bi-valuation $p(x|y) \equiv z(x,y)$ that follows a \emph{sum
rule}, a \emph{product rule} and a \emph{Bayes' theorem}. This
bi-valuation is a measure that quantifies the degree to which one
statement implies another, and is essentially the degree of
implication that Cox considered in his seminal work
\cite{Cox:1946,Cox:1961}. Thus order theory gives rise to
probability theory \cite{Knuth:laws,Knuth:Neurocomp-preprint}.

Probability theory, however, does not instruct us on how to assign
priors (which can be considered as valuations, eg. $p(x|\top)
\equiv v(x)$). An important theorem by Gian-Carlo Rota
\cite[Theorem 1, Corollary 2, p.35]{Rota:combinatorics} makes this
fact clear:

\begin{theorem}
A valuation $v$ in a finite distributive lattice $L$ is uniquely
determined by the values it takes on the set of join-irreducibles
of $L$, and these values can be arbitrarily assigned.
\end{theorem}

There is no information in the Boolean algebra, and hence the
inferential calculus (probability theory), to instruct us in
assigning priors.  We must instead rely on additional principles,
such as symmetry, constraints, and consistency with other aspects
of the problem to assign priors.  However, once the priors are
assigned, order-theoretic principles dictate the remaining
probabilities through the inferential calculus.

\section{Relevancy}
Cox defined a question as the set of all possible logical
statements that answer it \cite{Cox:1979}. With questions being
described by sets, their natural algebra is the distributive
algebra, with the join $\vee$ and meet $\wedge$ identified with
the set union $\cup$ and set intersection $\cap$, respectively.
The natural ordering relation among questions is the relation
`\textit{answers}', which can be represented mathematically by
$\subseteq$. Thus questions possess two algebraic operations
$\vee$ and $\wedge$ analogous to the familiar disjunction and
conjunction of logical statements.  In fact, we even use the words
`or' and `and' to describe them in spoken language. However, the
algebra is not Boolean as Cox surmised, since the definition of a
question is rather restrictive (i.e. not all sets of logical
statements correspond to questions). Thus questions do not, in
general, have complements \cite{Knuth:Questions}.

In order theory, Cox's definition of a question is equivalent to
saying that a question is a down-set, where a down-set is the set
of all poset elements that contain every element that includes any
other element of the set
\cite{Davey&Priestley,Knuth:Neurocomp-preprint}
\begin{definition}[Down-set]
A \textit{down-set} is a subset $J$ of an ordered set $L$, written
$J = \downset L$, where if $a \in J$, $x \in L$, $x \leq a$ then
$x \in J$.
\end{definition}
where $J$ is the question, $L$ is the ordered set of logical
statements, and $\leq$ is `\textit{implies}' $\rightarrow$. The
question lattice is then formed by taking the set of down-sets of
the assertion lattice and ordering them according to $\subseteq$.
This operation, called \textit{the ordered set of down-sets}
$\EuScript{O}$, takes the assertion lattice to the question
lattice, $\EuScript{Q} = \EuScript{O}(\EuScript{A})$. Figure
\ref{fig:fig1} shows the lattice $\EuScript{Q}_3$ generated from
$\EuScript{A}_3$. The lattice $\EuScript{Q}_3$ depicts all
possible questions that can be asked in this example. Note that $A
\equiv \downset a$, $AN \equiv \downset a \vee n$, $AKN \equiv
\downset a \vee k \vee n$, and $AN \vee AK \equiv AN \cup AK$. The
lattice $\EuScript{A}$ is dual to the lattice $\EuScript{Q}$ in
the sense of \textit{Birkhoff's Representation Theorem}
\cite{Davey&Priestley,Knuth:Questions}, which relates a
distributive lattice to its ordered set of down-sets.

The ideal questions $\EuScript{I}$, are the set of
join-irreducible elements of $\EuScript{Q}$. They are not
practical questions, but are useful mathematical constructs since
they are isomorphic to the assertion lattice. The real questions
$\EuScript{R}$ are the set of all questions that can be answered
by each of the atomic statements $a$, $k$, or $n$. The question $I
= A \vee K \vee N$ is a special real question that I call the
\emph{central issue} \cite{Knuth:Neurocomp-preprint}. It is the
unique real question that answers all the others.  In this
example, it asks `Precisely who stole the tarts?'

Just as in the lattice of assertions, we can define the degree to
which one question answers another by $d(X|Y) = z(X,Y)$. Since the
lattice of questions is distributive, there exists a sum rule, a
product rule, and a Bayes' theorem. This degree is called
relevance, and due to the duality between $\EuScript{A}$ and
$\EuScript{Q}$ it is entirely reasonable to expect that relevance
on $\EuScript{Q}$ is related to probability on $\EuScript{A}$. We
explore this in the next section.

\section{Consistency between Probability and Relevancy}
The sum, product and Bayes' rules ensure consistency \emph{within}
the assertion and question lattices, however our assignments of
probabilities and relevances must also be mutually consistent with
one another. Rota's theorem assures that we need only to determine
the relevances of the join-irreducible questions; the rest follow
from the inquiry calculus.

In this section, I show that the form of the relevance is uniquely
determined by requiring consistency between the probability
measure defined on the assertion lattice $\EuScript{A}$ and the
relevance measure defined on its isomorphic counterpart, the
lattice of ideal questions $\EuScript{I}$. This demonstration
requires but a single assumption: the degree to which the top
question $\top$ answers an ideal question $X$ depends only on the
probability of the assertion $x$ from which the question $X$
originated. That is, given the ideal question $X = \downset x$
\begin{equation}
d(X | \top) = H(p(x|\top)), \label{eq:relevance-of-ideal}
\end{equation}
where $H$ is a function to be determined.

There are four important consistency requirements imposed by the
lattice structure and the induced calculus. First, the sum rule
(\ref{eq:incl-excl}) for questions demands that given three
questions $X, Y, Q \in \EuScript{Q}$ the relevance is
\emph{additive} only when $X \wedge Y = \bot$
\begin{equation}
d(X \vee Y | Q) = d(X | Q) + d(Y | Q), \quad \mbox{iff} \quad X
\wedge Y = \bot. \label{eq:additivity}
\mbox{~~~~~~~~~~~~~~~(\emph{additivity})}
\end{equation}
and is \emph{subadditive}
\begin{equation}
d(X \vee Y | Q) \leq d(X | Q) + d(Y | Q). \label{eq:subadditivity}
\mbox{~~~~~~~~~~~~~~~~~~~~~~~~~~~~~~~~~~~~~~~(\emph{subadditivity})}
\end{equation}
in general; a result of the terms in the sum rule
(\ref{eq:incl-excl}), which avoid double-counting the overlap
between the two questions
\cite{Knuth:laws,Knuth:Neurocomp-preprint}. Commutativity of the
join requires that
\begin{equation}
d(X_1 \vee X_2 \vee \cdots \vee X_n | Q) = d(X_{\pi(1)} \vee
X_{\pi(2)} \vee \cdots \vee X_{\pi(n)} | Q) \label{eq:symmetry}
\mbox{~~~~~(\emph{symmetry})}
\end{equation}
for all permutations $(\pi(1), \pi(2) \cdots, \pi(n))$ of $(1, 2,
\cdots, n)$. Thus the relevance must be \emph{symmetric} with
respect to the order of the joins.

Last, since any assertion $f$, known to be false can be identified
with the bottom $\bot$ in $\EuScript{A}$, its corresponding ideal
question $F = \downset f \in \EuScript{I}$ can be identified with
$\bot$ in $\EuScript{Q}$. Since for all questions $X \in
\EuScript{Q}$ it is true that $X \vee \bot = X$, we have the
\emph{expansibility} condition
\begin{equation}
d(X_1 \vee X_2 \vee \cdots \vee X_n \vee F | Q) = d(X_1 \vee X_2
\vee \cdots \vee X_n | Q). \label{eq:expansibility}
\mbox{~~~~~~~~~~~(\emph{expansibility})}
\end{equation}

I now define a \emph{partition question} as a real question where
its set of answers are neatly partitioned. More specifically
\begin{definition}[Partition Question]
A \textit{partition question} is a real question $P \in
\EuScript{R}$ formed from the join of a set of ideal questions $P
= \bigvee_{i=1}^{n} X_i$ where $\forall\; X_j, X_k \in
\EuScript{J}(\EuScript{Q})$, $X_j \wedge X_k = \bot$ when $j \neq
k$.
\end{definition}

For a partition question $P$, the degree to which the top question
$\top$ answers $P$ can be easily written using
(\ref{eq:additivity})
\begin{equation}
d(P | \top) = d(\bigvee_{i=1}^{n} X_i | \top) =
\sum_{i=1}^n{H(p(x_i|\top))}.
\end{equation}
An important result from Acz\'{e}l et al.
\cite{Aczel+etal:natural} states that if a function of this form
satisfies additivity (\ref{eq:additivity}), subadditivity
(\ref{eq:subadditivity}), symmetry (\ref{eq:symmetry}), and
expansibility (\ref{eq:expansibility}), then the unique form of
the function is a linear combination of the Shannon and Hartley
entropies
\begin{equation}
d(P | \top) = a \;H_m(p_1, p_2, \cdots, p_n) + b \;{}_{o}H_m(p_1,
p_2, \cdots, p_n), \label{eq:general-result}
\end{equation}
where $p_i \equiv p(x_i|\top)$, $a, b$ are arbitrary non-negative
constants. The Shannon entropy \cite{Shannon&Weaver} is defined as
\begin{equation}
H_m(p_1, p_2, \cdots, p_n) = - \sum^n_{i=1}{p_i \log_2 p_i},
\label{eq:shannon}
\end{equation}
and the Hartley entropy \cite{Hartley} is defined as
\begin{equation}
{}_{o}H_m(p_1, p_2, \cdots, p_n) = \log_2 N(P), \label{eq:hartley}
\end{equation}
where $N(P)$ is the number of non-zero arguments $p_i$. An
additional condition suggested by Acz\'{e}l states that the
Shannon entropy is the unique solution if the result is to be
small for small probabilities \cite{Aczel+etal:natural}; that is,
the relevance varies continuously as a function of the
probability. This result is important since it rules out the use
of other types of entropy, such as the Renyi and Tsallis
entropies, for the purposes of inference and inquiry.

Given these results, the relevance of an ideal question
(\ref{eq:relevance-of-ideal}) can be written as
\begin{equation}
d(X | \top) = -a p(x | \top) \log_2 p(x | \top),
\end{equation}
which is proportional to the probability-weighted surprise. The
sum rule allows us to calculate more complex relevances, such as
that of the central issue
\begin{equation}
d(A \vee K \vee N | \top) \propto - p_a \log_2 p_a - p_k \log_2
p_k - p_n \log_2 p_n, \label{eq:relevance-of-T}
\end{equation}
where $p_a \equiv p(a | \top), \cdots$, and we have set the
arbitrary constant $a = 1$.

With the relevances of the join-irreducible questions defined, the
inquiry calculus allows us to compute the relevance between any
two questions. The degree to which an arbitrary question $Q$
answers a question $X$ can be found from $d(X | \top)$ by
recognizing that $d(X | Q) = d(X | Q \wedge \top)$ and using
Bayes' Theorem. Furthermore, the relevance of $Q$ to the join of
two questions such as $AN \cup KN \equiv AN \vee KN$ is
\begin{equation}
d(AN \vee KN | Q) = d(AN | Q) + d(KN | Q) - d(AN \wedge KN | Q),
\end{equation}
which is clearly related to the mutual information, although the
conditionality of this measure absent in the information-theoretic
notation. Thus the relevance of the join of two questions is akin
to mutual information, which describes what the two questions ask
in common. Similarly, the relevance of the meet of two questions
$d(AN \wedge KN | Q)$ is akin to the joint entropy. In the context
of information theory, Cox's choice in naming the common question
and joint question is very satisfying.

The inquiry calculus holds new possibilities. Not only does it
allow for conditionality, which is obscured and implicit in
information theory, but the relevance of questions comprised of
the joins of multiple questions can be computed using the sum
rule, which proves to be the \emph{generalized entropy}
conjectured by Cox \cite{Cox:1961,Cox:1979}. Furthermore, special
cases of these relevances have appeared before in the literature
\cite{Knuth:Neurocomp-preprint}. The result presented here is a
well-founded generalization of information theory, where the
relationships among a set of any number of questions can be
quantified.

Last, it should be noted that by setting $a=b$ in
(\ref{eq:general-result}), and using (\ref{eq:shannon}),
(\ref{eq:hartley}), we get
\begin{equation}
H_m(p_1, p_2, \cdots, p_n) = - \sum^n_{i=1}{p_i \log_2
\frac{p_i}{\frac{1}{n}}},
\end{equation}
which is the relative entropy based on a uniform measure.

\section{Maximum Entropy}
This result provides new insights into the assumptions underlying
the \emph{Principle of Maximum Entropy}
\cite{Jaynes:MaxEnt,Jaynes:Book}. Consider both the assertion
lattice $\EuScript{A}$ and its dual the question lattice
$\EuScript{Q}$. What does it mean to assign probabilities to
$\EuScript{A}$ by maximizing the entropy? When we `maximize the
entropy', we are actually maximizing the relevance of the top
question $\top$ to the central issue $I$, i.e. we maximize
$d(I|\top)$. This says that we are setting up the probability
assignments so that the question that asks everything is maximally
relevant to the central issue. To understand what this means, it
is useful to see what happens in a special case. In the situation
where we have no constraints, this results in assigning uniform
prior probabilities to the join-irreducible elements of
$\EuScript{A}$. What if in the case of three statements, we assign
the probabilities non-uniformly: $p(x_1|\top) = 0.1$, $p(x_2|\top)
= 0.4$, $p(x_3|\top) = 0.5$. In this case, the central issue no
longer has the maximal relevance. Instead, the question defined by
the set $X_1X_2 \vee X_3 = \{x_1 \vee x_2, x_1, x_2, x_3, \bot\}$
has the maximal relevance. This suggests that a re-parametrization
of the problem is more \emph{relevant} $\{x_1 \vee x_2, x_3\}$.
Thus when we assign priors based on maximizing the entropy, we are
relying on the fact that we believe that we have a relevant
parametrization of the problem. In other words: we have identified
the \emph{relevant variables}.

\section{Discussion}
Within the last few years there has a been a surge of interest in
Bayesian methods.  Much of this is due to the fact that Bayesian
methods work, and work well.  However, with this surge of activity
the ideas of Jaynes and Cox are slowly being lost as converted
statisticians focus more and more on mathematical rigor and less
on the basic concepts. Ironically, it is this focus on
mathematical rigor and loss of the basic concepts that buried
Bayesian methods in the 19th century. Cox's realization that
Bayesian probability theory is the only theory consistent with
Boolean logic is key, since it rules out all other possible
theories of inference. Jaynes' realization that the entropy of
statistical mechanics is related to Shannon's entropy, and that
one can use it to assign priors is crucial since it ties together
the physics of thermodynamics to inference. However, the
successful application of inference in several key areas of
physics seems to be of little interest to statisticians, which is
puzzling given both the great success of the theories and the
great mysteries that they simultaneously resolve and reveal.

The basic concepts are key, because it is by fully understanding
these concepts that we can generalize these ideas to form new
theories.  I have found that Cox's idea of introducing a real
number representing a degree of belief can be generalized to
introducing a real-valued function generalizing the zeta function
of a lattice.  This allows one to take any ordered set that forms
a lattice and introduce a measure describing the degree of
inclusion associated with that ordering relation.

Cox's definition of a question is the definition of a down-set in
order theory. With this definition in hand, I showed that the
ordered set of down-sets of assertions gives rise to the set of
all possible questions, which forms a distributive lattice.
Realizing that Caticha's results on quantum mechanical
experimental setups demonstrate that sum and product rules are
associated with distributive lattices, I showed that the calculus
of inquiry has sum and product rules analogous to the inferential
calculus.  This paper extends these results by requiring
consistency between the measures assigned to the lattice of
assertions and the lattice of questions.  With yet another
important result by Acz\'{e}l, I have shown that the relevance
measure on the lattice of questions is based on the Shannon
entropy.  This is significant since it rules out the use of other
entropies in inference (eg. Renyi entropy and Tsallis entropy), as
well as inquiry. The result is that the inquiry calculus and the
relevance measure is a natural generalization of information
theory. Furthermore, these results provide a new perspective on
the role of Maximum Entropy in prior probability assignment.

\begin{theacknowledgments}
This work supported by the NASA IDU/IS/CICT Program. I am deeply
indebted to Ariel Caticha, Bob Fry, Carlos Rodr\'{i}guez, Janos
Acz\'{e}l, Ray Smith, Myron Tribus, David Hestenes, Larry
Bretthorst, Jeffrey Jewell, and Bernd Fischer for insightful and
inspiring discussions, and many invaluable remarks and comments.
\end{theacknowledgments}


\doingARLO[\bibliographystyle{aipproc}]
          {\ifthenelse{\equal{\AIPcitestyleselect}{num}}
             {\bibliographystyle{arlonum}}
             {\bibliographystyle{arlobib}}
          }
\bibliography{knuth}

\end{document}